\documentclass[pra,preprint,twocolumn]{revtex4}
\usepackage{amsfonts}
\usepackage{amsmath}
\usepackage{graphicx}

\begin{document}

\title{Proper and improper density matrices and the consistency of the
Deutsch model for Local Closed Timelike Curves}
\author{Augusto C\'{e}sar Lobo}
\affiliation{Physics Department, Federal University of Ouro Preto}
\author{Ismael Lucas de Paiva}
\affiliation{Physics Department, Federal University of Minas Gerais}
\author{Pedro Ruas Dieguez}
\affiliation{Physics Department, Federal University of Minas Gerais}


\begin{abstract}
We discuss the concept of proper and improper density matrixes and we argue
that this issue is of fundamental importance for the understanding of the
quantum-mechanical CTC model proposed by Deutsch. We arrive at the
conclusion that under a realistic interpretation, the distinction between
proper and improper density operators is not a relativistically covariant
notion and this fact leads to the conclusion that the D-CTC model is
physically inconsistent.
\end{abstract}

\maketitle

\section{Introduction}

Though quantum mechanics has been indisputably recognized for a long time
now as\ much as a counter-intuitive theory as it is an extraordinarily
effective one - it may be said that the issue of its ``rigidity" as a
physical model has arisen only in more recent times \cite{weinbergdreams}.
The ``rigidity" of a theory can be thought as a qualitative measure of how
much the entire theoretical construction is somehow constrained by its
principles in the sense that by tampering with one of them would\ ``bring
down" the whole structure, making the theory incoherent or inconsistent.

In 1989, Weinberg \cite{weinberg1989-1,weinberg1989-2} initiated an
investigation on how far one could modify ordinary quantum mechanical
principles by adding some extent of non-linearity to its axioms and
surprisingly it turned out to be a very difficult enterprise. Almost
immediately it was pointed out that this would lead to superluminal
communication or \textit{signaling} \cite{Gisin,Polchinski}. As is
well known, this leads to paradoxes and is considered by most physicists to
be nonphysical.

In fact, in 1949, G\"{o}del found a solution of general relativistic field
equations that exhibits \textit{closed time-like curves} (CTCs) \cite{Godel}%
. Most physicists at the time (including Einstein) found this result quite
disturbing, but disregarded it as probably nonphysical, preferring to
believe maybe that a future more complete understanding of the physics
involved should somehow rule out such kind of a solution.

Yet, in 1991, Deutsch presented a computational analysis of the problem --
both classical and quantum mechanical \cite{Deutsch}. He addressed the
inherent free-will kind of problems that stems from grandfather-like
paradoxes in a more technical and less anthropomorphic way than usual. He
assumed the existence of a region of space-time (CTC) which violates the
usual chronological respecting space-time (CR) and considered a
computational circuit where one or more bits or qubits may enter the CTC and
travel back in time meeting itself at an earlier moment. He also conceived
that a CR\ system may interact with the CTC system by some unspecified
unitary interaction. He showed that, in general, classical computational
logic is inconsistent with the existence of CTCs in the sense that \textit{%
not all} possible input states have a consistent solution -- this is the
technical way of stating the free-will problem. Yet, he suggested a quantum
mechanical model for systems (allowing for mixture states), that exhibits
output solutions for \textit{every} input state, seemingly circumventing the
paradoxes. Let $\hat{\rho}^{(i)}$ be the initial state of the chronology
respecting system and $\hat{\rho}_{CTC}$ the state captured in the CTC loop,
then he proposed the following self consistent relation that $\hat{\rho}%
_{CTC}$ must obey: 
\begin{equation}
\hat{\rho}_{CTC}=tr_{CR}[\hat{U}(\hat{\rho}^{(i)}\otimes \hat{\rho}_{CTC})%
\hat{U}^{\dag }]  \label{self-consistency relation for rho CTC}
\end{equation}

The physical meaning of the above equation is clear. The CR system (Alice)
comes close enough to the CTC system (Bob), interacts with it in a limited
region of space-time and becomes part of a larger entangled state $\hat{U}(%
\hat{\rho}^{(i)}\otimes \hat{\rho}_{CTC})\hat{U}^{\dag }$ under the
interaction modeled by the global unitary operator $\hat{U}$. After a while,
the system moves away and Bob's system is obtained by partial tracing out
Alice's system. The above equation is a way to impose that the \textit{output%
} state is the \textit{same} as the \textit{input} one, obeying the CTC
criterion with \textit{no paradoxes} as Deutsch showed that there is always 
\textit{at least one} solution (there may be more than one.) Note that $\hat{%
\rho}_{CTC}$ depends on the initial state $\hat{\rho}^{(i)}$ and on the
unitary operator $\hat{U}$ (the interaction). Alice's system will have
changed to 
\begin{equation}
\hat{\rho}^{(f)}=tr_{CTC}[\hat{U}(\hat{\rho}^{(i)}\otimes \hat{\rho}_{CTC})%
\hat{U}^{\dag }]  \label{equacao da dinamica do sistema CR}
\end{equation}

The two equations above clearly imply a \textit{non-linear} evolution for
Alice's system because (\ref{equacao da dinamica do sistema CR}) means that $%
\hat{\rho}^{(f)}$ depends both on $\hat{\rho}^{(i)}$ and $\hat{\rho}_{CTC}$,
(and on the interaction) but the latter also depends on $\hat{\rho}^{(i)}$.
This feature is a novel ingredient that goes beyond the linear evolution
described by Schr\"{o}dinger's equation. Since Deutsch's proposal, many
results appeared in the literature where it was argued that quantum
mechanics together with Deutsch's model for closed time-like curves (D-CTC)
is more powerful than ordinary quantum physics. Claims as cloning of quantum
states \cite{Ahn2010}, solving NP-complete problems in polynomial time \cite%
{Brun2003} and distinguishing non-orthogonal states \cite{Brun} have been
reported.

In this paper we discuss the fact that under a certain interpretation of
quantum mechanics, the ability to distinguish non-orthogonal states leads
directly to inconsistencies in the Deutsch protocol for quantum systems
traversing CTCs. We argue that the concept of a density matrix to be proper
or improper is a relativistically non-covariant notion and that this result
leads us to the above conclusion after we examine in detail a specific
instance of a protocol for non-orthogonal state discrimination presented by
Brun \textit{et al}.

The paper is structured as follows: In Section II, we will briefly review
the non-local properties of a EPR-like pair of qubits and discuss why by the
ordinary interpretation of quantum mechanics, the non-distinguishability of
non-orthonormal states protects the theory from superluminal communication
despite the non-local information contained in the entangled states. In
Section III, we discuss some interpretational issues on the difference
between proper and improper density matrices in quantum mechanics that are
crucial to our main conclusions. In Section IV, we briefly review the
results in \cite{Brun} that allows the discrimination of non-orthogonal
states with the Deutsch model and how this implies that the D-CTC model is
not consistent. In Section V, we conclude our work with a discussion about
the physical meaning of these results and we also set stage for further work.

\section{Non-discrimination of non-orthogonal states in ordinary QM}

The celebrated 1935 EPR\ paper showed, for the first time, what Einstein
coined as \textquotedblleft spooky action at distance" of Bell-like states
even though it was early recognized that it was impossible to use entangled
states for superluminal communication \cite{EPR}. Suppose two parties, Alice
and Bob meet to establish an interaction between their qubits to create a
maximal entangled pair of qubits in the anti-correlated form 
\begin{equation}
\left\vert \Psi \right\rangle =\frac{1}{\sqrt{2}}\left( \left\vert
z+\right\rangle \otimes \left\vert z-\right\rangle -\left\vert
z-\right\rangle \otimes \left\vert z+\right\rangle \right)  \label{EPR pair}
\end{equation}%
After this, they go apart and each one has access only to his own qubit.
Both parts will describe their own system by the density matrix

\begin{equation}
\hat{\rho}_{Alice}=\hat{\rho}_{Bob}=\frac{1}{2}\hat{I}
\label{maximally entangled mixed qubit}
\end{equation}

This means that (\ref{EPR pair}) can be written as 
\begin{equation}
\left\vert \Psi \right\rangle =\frac{1}{\sqrt{2}}\left( \left\vert \hat{n}%
\right\rangle \otimes \left\vert -\hat{n}\right\rangle -\left\vert -\hat{n}%
\right\rangle \otimes \left\vert \hat{n}\right\rangle \right)
\label{Bell State}
\end{equation}

Where the state $\left\vert \hat{n}\left( \theta ,\varphi \right)
\right\rangle $ defined as 
\begin{equation}
\left\vert \hat{n}\left( \theta ,\varphi \right) \right\rangle =\cos \left(
\theta /2\right) \left\vert z+\right\rangle +e^{i\varphi }\sin \left( \theta
/2\right) \left\vert z-\right\rangle
\end{equation}%
is projected to an arbitrary point with $\left( \theta ,\varphi \right) $
coordinates on the surface of the Bloch sphere. Note that Hilbert space
orthonormality of two states means that the vectors are projected to
antipodal points on the Bloch sphere. To convey superluminal information,
one part should be able to distinguish \textit{different directions} in the
Bloch sphere geometry, which means the ability to \textit{discriminate} 
\textit{non-orthogonal} states.

Thus, Alice may \textquotedblleft collapse\textquotedblright\ the mixed
state by measuring her state in \textit{any direction} of the Bloch Sphere,
transforming non-locally Bob's description to a pure state. But since we may
assume that Bob is far away, he has no way to find out about Alice's
measurement without receiving a message from her. Alice and Bob could 
\textit{agree before hand} to measure only over \textit{two different}
directions, let us say the $x$ and $z$ directions. Alice could hopelessly
try to code one classical bit of information in these distinguished
directions, by assigning logical values to them, for instance, $0$ to the $z$
direction and $1$ to the $x$ direction. By choosing which direction to
measure her qubit, she would collapse the whole system instantaneously (in
some inertial reference system).

Suppose she chooses to send the $0$ bit and after her $z$ measurement, she
\textquotedblleft collapses\textquotedblright\ her state to $\left\vert
z+\right\rangle $. There is no way that Bob can receive this classical piece
of information. If he chooses to measure the $z$ direction, he will
certainly obtain $\left\vert z-\right\rangle $, but there is no way for him
to know if his collapsed pure state was in the $x$ or $z$ directions without
receiving information from Alice in the usual (subluminal) manner. The
intrinsic randomness of quantum mechanics avoids superluminal communication.
\begin{figure}[!htb]
\begin{center}
     \includegraphics[scale=.28]{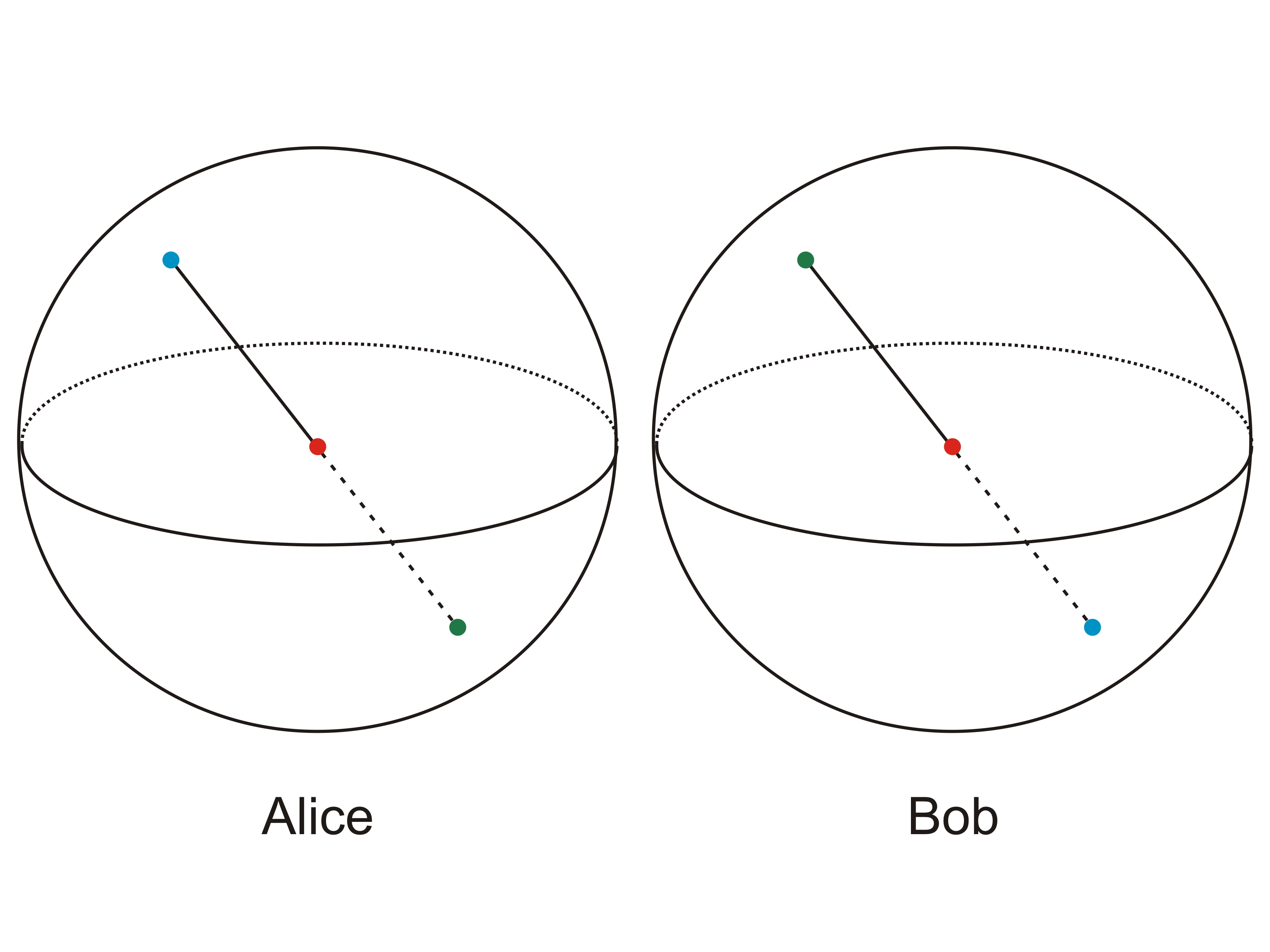}
\end{center}
\end{figure}

By the other way around, the ability of discrimination of non-orthogonal
states leads necessarily to signaling as can be easily seen. Most papers on
no-signaling and quantum mechanics discuss this issue in terms of the
no-cloning theorem (\cite{westmoreland1998quantum,scarani2005quantum, pawlowski2009new}) because both these abilities are actually
equivalent \cite{caves1996quantum}. Curiously, we were incapable of finding
in the literature any explicit discussion of the no-signaling property of
quantum mechanics in terms of the indistinguishability of non-orthogonal
states. This issue seems to have solely been discussed in terms of the
no-cloning theorem. Maybe this is because signaling is such an obvious
consequence of non-orthogonal state distinguishability that most authors
just take it for granted. On possible exception is Gisin's 1989 paper where
he refutes Weinberg's attempt to construct a non-linear quantum mechanical theory 
\cite{Gisin}. Indeed in his paper, he manages to ultimately distinguish
non-orthogonal states, but only after a quite elaborate construction where
there must be a third party besides Alice and Bob that must continuously
provide a stream of entangled qubits, each one sent to one of the other
parts. (Alice and Bob). We show in our above analysis that any ``non-unitary
machine" capable of discriminating non-orthogonal states immediately implies
in superluminality, a result that seems quite obvious, but which seems that
has not been explicitly stated anywhere before in the literature.

In fact, going back to the EPR protocol discussed above, if Bob has access
to some device capable of discriminating between the four state vectors of
their common alphabet, Alice will clearly be able to communicate a classical
bit in a superluminal way by choosing in which of the two agreed directions
she performs her measurement.

\section{Some interpretational issues of Quantum Mechanics}

\subsection{The interpretation of the state vector}

The picture of entanglement presented in the last Section goes along with
the most usual interpretation of quantum physics. This interpretation (let
us agree to call it a \textit{realistic} interpretation) views the
entanglement phenomenon as exhibiting genuine non-local properties even if
this non-locality does \textit{not} lead to signaling. In this
interpretation, the entangled state (\ref{Bell State}) can be seen as a
non-causal channel connecting two distant observers such that if a pair of
measurements are performed by Alice and Bob (one measurement each - at
space-like separated events) then it follows that there is no consistent way
to assign any causal relation between them.

One may say that in this view, one thinks of the state vector (or its
projection onto the space of rays) as a \textit{real} objective property of
the system. These quantum channels are behind many of the modern
applications of quantum information as quantum teleportation and
cryptography.

A very different view is taken by some physicists like \cite{Peres2004} or 
\cite{wallman} for example. This approach to quantum mechanics may be
described as \textit{epistemic} in the sense that one assigns to the state
vector the subjective property of describing only the \textit{knowledge} of
an observer about the system. For those who subscribe to this
interpretation, there is no such thing as non-local phenomena in quantum
mechanics, because the non-local collapse of the global state vector is a
non-physical process. In \cite{wallman}, for instance, the authors conclude
that the Deutsch scheme manages only to \textit{conceal} the paradoxes
involved in time-travelling instead of resolving them.

Yet, regardless of these different interpretations, for linear quantum
mechanics, it is noncontroversial that one arrives at the same physical
predictions. One could wonder if these subtle philosophical distinctions are
then actually physically relevant. But something quantum mechanics has
taught us during the last century is that one should be specially careful
before jumping to such a conclusion. After all, the Bohr-Einstein debate was
generally considered (for three decades) as being of a philosophical
character before Bell introduced in 1964 his inequalities for local hidden
variable correlations and showed that the dispute was physically verifiable.
One may even make the case that this was the moment when the embryo of
modern quantum information science was laid. Our opinion is that indeed for
linear quantum mechanics, there may not be any physical differences between
these two approaches. But if any non-linearity is introduced as some
physicists expect for a full consistent quantum theory of gravity (see \cite%
{Hawking2005} and \cite{Penrose1994} for opposite argumentations on this
issue), then the situation will probably be very different as we argue in
the next sections.

\subsection{The interpretation of the density matrix}

There are two very different approaches to the density matrix concept. The
first may be called the text-book concept of a density matrix and it is a
reminiscent of the historical way that statistical concepts were introduced
in classical physics. One is given a large ensemble of $N$ identical quantum
systems and one supposes that the ensemble can be partitioned into many
sub-ensembles labeled by an index $\alpha =1,2,...m$ where $m$ is the number
of different sub-ensembles $\alpha $ characterized by the fact that every
quantum system that belongs to it is in a pure state $\left\vert \psi
_{\alpha }\right\rangle $. Now suppose further that an observer may ``pick"
one system from the ensemble in a \textit{random} way. Notice that this
randomness is purely ``classical" in some sense. The probability that the
observer picks up a system from a subensemble $\alpha $ is clearly $%
P_{\alpha }=n_{\alpha }/N$ where $n_{\alpha }$ is the number of systems in
sub-ensemble $\alpha $. Of course it clearly holds that the probability $%
P_{\alpha }$ is automatically normalized as it indeed must be. Given an
arbitrary observable $\hat{O}$, its expectation value for $\left\vert \psi
_{\alpha }\right\rangle $ is $\left\langle \psi _{\alpha }\right\vert \hat{O}%
\left\vert \psi _{\alpha }\right\rangle $ and if the observer repeats the
procedure many times, the average expectation value is clearly 
\begin{equation*}
\sum_{\alpha }P_{\alpha }\left\langle \psi _{\alpha }\right\vert 
\hat{O}\left\vert \psi _{\alpha }\right\rangle =tr\left( \hat{\rho}\hat{O}%
\right) 
\end{equation*}%
where 
\begin{equation*}
\hat{\rho}=\sum_{\alpha }P_{\alpha }\left\vert \psi _{\alpha
}\right\rangle \left\langle \psi _{\alpha }\right\vert 
\end{equation*}%
is defined as the density matrix. This definition is easily seen as an
epistemic definition where one considers $P_{\alpha }$ a classical
probability distribution that is subjective in the sense that it reveals the 
``classical ignorance" of the observer about which pure state $\left\vert
\psi _{\alpha }\right\rangle $ the system actually belongs to. This
epistemic definition is known also as a \textit{proper} density matrix \cite{despagnat76}.

A second conception of density matrix has a purely quantum mechanical
origin. Given an entangled state $\left\vert \Psi \right\rangle \in
W=W^{(a)}\otimes W^{(b)}$ of two subsystems (Alice and Bob's subsystems),
suppose (after the global entangled state has been created locally) each one
has physical access only to its own subsystem. Let $\hat{O}$ be again an
arbitrary observable of Alice's subsystem, then she can define a density
matrix $\hat{\rho}_{\left\vert \Psi \right\rangle }$ by the following
equation%
\begin{equation*}
\left\langle \Psi \right\vert \hat{O}\otimes \hat{I}\left\vert \Psi
\right\rangle =tr\left( \hat{\rho}_{\left\vert \Psi \right\rangle }\hat{O}%
\right) \qquad \text{(for all\ }\hat{O}\text{)}
\end{equation*}

The above equation defines a mapping from the space of rays (the projective
space of the full quantum space $W$) to the space of linear operators in $%
W^{(a)}$. This (non-linear) mapping is called the \textit{partial trace} and
it can be conducted in the same manner for Bob's subspace. This definition
of density matrix is known as an \textit{improper mixture} and it is
commonly thought as an \textit{ontological} description of a mixed state in
some interpretations and should then be seen as essentially different from
the former case.

What is remarkable is that mathematically, both definitions lead to
equivalent descriptions. Both (proper and improper mixtures) are hermitian,
positive and unit trace operators. Some authors deny the existence of proper
mixtures in the sense that all density matrices can be thought as resulting
from a partial trace. There is some controversy on this matter. (See \cite{despagnat1998} and \cite{Anandan} for opposite opinions on this issue.) But
consider now the following two experiments:

\begin{enumerate}
\item Alice tosses a fair coin in her lab to decide if she produces a pure $%
\left\vert z+\right\rangle $ or $\left\vert z-\right\rangle $ state and then
sends the state to Bob.

\item Alice initially produces an entangled 2-qubit state as in (\ref{Bell
State}) and then measures one of the qubits in the $z$ direction and sends
the other qubit to Bob.
\end{enumerate}

It is clear that both preparations must lead to physically indistinguishable
states for Bob (at least for usual linear quantum mechanics) represented
both by the same proper density matrix given by (\ref{maximally entangled
mixed qubit}). If one introduces elements of non-linearity as is the case of
the D-CTC prescription, things are not as simple as we shall see in the next
Section.

\section{Non-orthogonal state discrimination protocol with Deutsch CTC's}

In \cite{Brun}, Brun\textit{\ et al} exhibited a quantum computational
protocol with access to a Deutsch CTC that allows discrimination of
arbitrary non-orthogonal state vectors.

Let $C=\{\left\vert \psi _{k}\right\rangle \}$, ($k=1,...n$) be a set of $n$
non-orthogonal normalized state vectors belonging to a finite $n$%
-dimensional Hilbert space $W$ of the CR system. Suppose that the input
state is given by the following pure density matrix $\hat{\rho}_{CR}=\hat{\pi%
}_{\left\vert \psi _{s}\right\rangle }=\left\vert \psi _{s}\right\rangle
\left\langle \psi _{s}\right\vert $ chosen from $C$. The protocol starts by
applying the \textit{swap operator} on the $W\otimes W_{CTC}$ system
followed by the controlled $\hat{U}$ operator 
\begin{equation}
\hat{U}=\sum_{j}\hat{\pi}_{j}\otimes \hat{O}_{j}  \label{controlled U}
\end{equation}%
where $\{\hat{\pi}_{j}\}$ is a family of one-dimensional projection
operators over an orthonormal basis $\{\left\vert u_{j}\right\rangle \}$ of $%
W$ and $\{\hat{O}_{j}\}$ is a family of unitary operators such that $\hat{O}%
_{k}\left\vert \psi _{k}\right\rangle =\left\vert u_{k}\right\rangle $. Brun%
\textit{\ et al} have shown that it is always possible to find a set $\{\hat{%
O}_{j}\}$ such that there is only \textit{one single} self-consistent
solution given by $\hat{\rho}_{CTC}=\hat{\pi}_{s}$. This allows the
existence of a bijective (non-linear) map between a set of non-orthogonal
states in $W$ and a set of orthogonal states $\{\left\vert
u_{j}\right\rangle \}$ which implies the discrimination of the elements of $%
C $.

With this result, one may approach again the EPR\ problem and conclude that
superluminal communication seems to be possible. In fact, suppose Alice and
Bob share a single pair of qubits in the Bell state given by (\ref{Bell
State}) where the first qubit stays in Alice's possession and the second
qubit travels with Bob. Suppose they have agreed previously over the code
described in Section I. Bob now must join his qubit with another qubit
system in a known state $\left\vert z+\right\rangle $ for instance. In this
way, after Alice measures her qubit in one of the two possible directions,
Bob's system will be in a state $\left\vert \xi _{j}\right\rangle \otimes
\left\vert z+\right\rangle $, with the state $\left\vert \xi
_{j}\right\rangle $ among the following four possibilities: $\left\vert \xi
_{0}\right\rangle =\left\vert z+\right\rangle $, $\left\vert \xi
_{1}\right\rangle =\left\vert z-\right\rangle $, $\left\vert \xi
_{2}\right\rangle =\left\vert x+\right\rangle $ or $\left\vert \xi
_{3}\right\rangle =\left\vert x-\right\rangle $. Bob then uses the D-CTC to 
\textit{distinguish} these non-orthogonal states. The input state is the
pure state $\left\vert \xi _{j}\right\rangle \otimes \left\vert
z+\right\rangle $ and the circuit will swap the CR\ and CTC\ system followed
by the controlled unitary operation (\ref{controlled U}) with $\left\vert
u_{0}\right\rangle =\left\vert z+\right\rangle \otimes \left\vert
z+\right\rangle $, $\left\vert u_{1}\right\rangle =\left\vert
z-\right\rangle \otimes \left\vert z+\right\rangle $, $\left\vert
u_{2}\right\rangle =\left\vert z+\right\rangle \otimes \left\vert
z-\right\rangle $, $\left\vert u_{3}\right\rangle =\left\vert
z-\right\rangle \otimes \left\vert z-\right\rangle $ and $\hat{O}_{i}\left(
\left\vert \xi _{i}\right\rangle \otimes \left\vert z+\right\rangle \right)
=\left\vert u_{i}\right\rangle $. In this way, Bob manages to know which
direction Alice decided to measure and her classical bit is communicated
with infinite speed.

In a recent paper \cite{Ralph}, the authors reach a different conclusion.
They show (as Deutsch assumes in his seminal paper) that quantum
correlations are lost in the D-CTC circuit by using a ``infinite loop"
equivalent circuit formulation where the problem is seen by the perspective
of the CTC system that is ``captured for eternity" in the loop. We subscribe
to this view but the authors then make a claim that we believe is mistaken:
they claim that this fact implies that superluminal phenomena cannot happen.
It seems that they believe that the ``collapse" of the global state of a
entangled state must happen only locally. But this belief is extraneous to
the usual view of quantum mechanics. And neither have we found elements in
Deutsch's original paper that supports such extreme notion. To subscribe to
this idea would be the same as introducing new elements in quantum physics
that indicate some kind of physical medium that would propagate the
``wave-function collapse" with finite speed. This is contrary to everything
we know about quantum mechanics including the well known results that
exhibit non-locality in the EPR\ experiments \cite{Aspect}.

Notice that Alice and Bob can arrange things such that for a certain moment
just \textit{before} Bob interacts with the D-CTC, Alice measures the system
in the way we described so that Bob is \textit{sure} that his state is pure.
His description would be given by a proper density matrix so that he can
treat his state in the same way that Brun\textit{\ et al} and Ralf\textit{\
et al} treat the states from the non-orthogonal alphabet of states chosen
between him and Alice. At this point - a physicist that believes in a
radically epistemic interpretation of quantum mechanics might reach a
different conclusion. He may say that there is no difference between proper
and improper mixtures and that they should be treated in the same way. This
is the ``linearity trap argument" put forward in \cite{Bennett} but
convincingly refuted in \cite{Cavalcanti} with a sound argument of
verifiability for any non-linear evolution. Indeed, Ralph \textit{et al}
also subscribe to Cavalcanti and Menicucci's opinion but seem not to
recognize that the same argument leads to signaling by the argument that we
present here.

An important point that must be made is that of taking seriously the fact
that the measurements are space-like separated events. This means that there
is a reference system for which Alice has not yet ``collapsed" the global
state and in this case indeed Bob will insert his qubit as an improper state
(\ref{maximally entangled mixed qubit}).

This result implies that the concept of proper and improper mixtures is 
\textit{not }a relativistically covariant notion. But one should note that
for usual linear quantum mechanics this is irrelevant because both concepts
are equivalent. In this case we have a completely different physical
picture. We agree that indeed in this scenario, the D-CTC should destroy all
quantum correlations and the output would again be the maximum entangled
state (\ref{maximally entangled mixed qubit}).

How can this be possible? The only conceivable answer to this question is
that it \textit{cannot} be. We summarize our analysis in the following way:
In one reference system, Bob's state is pure and he manages to receive
superluminal communication which ultimately leads to those same paradoxes
that Deutsch was originally trying to prevent and so it is then
inconsistent. In another reference state, his state is an improper mixture
and the signaling protocol fails. But this is inconsistent with the first
scenario. We conclude then\ that Deutsch's CTC\ model itself is \textit{%
overall} inconsistent.

\section{Concluding Remarks}

Does this analysis imply necessarily the conclusion that the Deutsch CTC
model is irreparably inconsistent after all? There are some fundamental
questions about the model that may be addressed. For instance, in Deutsch's
original scheme, the unitary operator to be applied in the interaction with
the CTC is supposed to be \textit{arbitrary}. Is this a reasonable
supposition on physical grounds? After all, shouldn't one expect that under
the very likely \textit{extreme} physical conditions that material particles
suffer under closed time-like curves as their world-lines, that the
physically possible unitary quantum evolutions should be constrained by the
rules of a (still unknown) consistent quantum theory of gravitation?

A full description of the largest class of unitary evolutions that \textit{%
not allow} arbitrary non-orthogonal state discrimination might turn out to
be rewarding in the sense that this information could be used as a clue for
what kind of interaction are or not permitted in such extreme scenarios.
Even if such a programme could be carried out successfully, the issue of the
relativistically noncovariance of the density matrix concept and the
resulting lack of inner consistency that it implies would also have to be
tackled -- see \cite{pati} for a recent discussion on related matters. Maybe
there is a manner of restraining the set of unitary operations that could at
the same time guarantee an equivalent physical description for all reference
frames. Of course, the fact that the many tasks that have been recently
claimed to be possible of implementation with Deutsch CTCs may be seen as a
sign that this model is inherently inconsistent after all.

It should also be noted that there is an alternative model to Deutsch's CTC.
The so called post-selection CTC (see \cite{Svetlichny}, \cite{Galvao}, \cite%
{1Lloyd2010}, \cite{2Lloyd2010} and \cite{brunwilde}) is a recent example of
a model that\textit{\ allows} non-orthogonal discrimination only for a set
of linearly independent states. Notice that this clearly forbids the
implementation of our signaling protocol between Alice and Bob because the
set $C$ of alphabet states is linearly dependent. This seems to imply that
the P-CTC model may indeed be more appropriate physically than the D-CTC
model. It is our intention to publish in the near future a paper
specifically on the P-CTC model in quantum mechanics.

That the ability to post-select may present counter-intuitive \textit{%
non-local time} properties for quantum mechanics has been noticed before 
\cite{Tollaksen}, \cite{aharonov2010}. It would not be surprising if
Aharonov's concepts of \textit{modular variables} and \textit{weak values}
turn out to play an essential role on this issue\ \cite{Aharonov1964}, \cite%
{Aharonov2007}. It is our opinion that the importance of research on this
topic is that of presenting toy models so that one may probe theoretically a
very elusive long-sought consistent quantum theory of gravity. Our work also
suggests that the construction of such a future theory may require that some
important foundational issues (as the intrinsic difference between proper
and improper density matrices) be addressed and adequately resolved first.

\begin{acknowledgments}
A.C.Lobo acknowledges financial support from NUPEC-Gorceix Foundation and
I.L.Paiva acknowledges financial support from FAPEMIG. All authors
acknowledge financial support from CNPq.
\end{acknowledgments}

\bibliographystyle{nature}
\bibliography{ctc}

\end{document}